\documentclass[letterpaper,12pt]{article}

\usepackage{authblk}
\usepackage{graphicx}
\usepackage{braket}
\usepackage{amsmath}
\usepackage{amssymb}
\usepackage{mathtools}
\usepackage{caption}
\usepackage{subcaption}
\usepackage{hyperref}
\usepackage{leftindex}
\usepackage[square,numbers,compress]{natbib}
\usepackage{color}

\usepackage[margin=1in,letterpaper]{geometry}
\usepackage{multirow}

\providecommand{\keywords}[1]
{
	\small	
	\textbf{\textit{Keywords---}} #1
}

\numberwithin{equation}{section}

\title{\vspace{-1.5cm}\textbf{\large Quantum Response of a Harmonically Trapped Detector to Classical and Non-classical Gravitational Fields}}
\author[1]{\normalsize Anom Trenggana \footnote{Electronic address: gstanomagung@gmail.com}}
\author[1,2]{\normalsize Freddy P. Zen \footnote{Electronic address: fpzen@fi.itb.ac.id}}
\author[3]{\normalsize Seramika Ariwahjoedi \footnote{Electronic address: sera001@brin.go.id}}
\affil[1]{\textit{\normalsize Theoretical Physics Laboratory, THEPi (Theoretical High Energy Physics) Division, Faculty of Mathematics and Natural Sciences, Institut Teknologi Bandung, Bandung 40132, West Java, Indonesia}}
\affil[2]{\textit{\normalsize Indonesian Center for Theoretical and Mathematical Physics (ICTMP), Institut Teknologi Bandung, Bandung 40132, West Java, Indonesia}}
\affil[3]{\textit{\normalsize Research Center for Quantum Physics, National Research and Innovation Agency (BRIN), South Tangerang 15314, Indonesia}}
\date{(\today)}

\begin{document}
\maketitle
\begin{abstract}
In this work, we study the response of a detector confined in a harmonic oscillator potential when interacting with classical and quantum gravitational fields. The detector response is characterized through transition probabilities between its energy levels, with the aim of investigating how non-classical properties of the gravitational field affect the detector dynamics. The quantum states of the gravitational field considered include coherent states and squeezed states. Our results show that the influence of the gravitational field on the detector transition probabilities is encoded in the two-time correlation function of the field. For coherent states, the structure of this two-time correlation function can be reproduced by an appropriately modeled classical gravitational field, particularly when the classical field is stationary. In contrast, for squeezed states, the two-time correlation function contains additional contributions that cannot be replicated within a classical description when the classical field is stationary, leading to a non-linear time dependence of the detector transition probabilities.
\end{abstract}


\newpage

\tableofcontents

\section{Introduction}\label{sec1}
The question of whether the gravitational field can be treated as a quantum field has, to date, not yet received a conclusive answer. This issue gained renewed prominence following the success of the LIGO detectors in observing gravitational waves for the first time in 2016 \cite{LIGO}. That achievement marked the beginning of a new era in gravitational-wave astronomy and revived fundamental questions concerning the quantum nature of the gravitational field. Since then, a substantial body of research has been devoted to addressing whether gravity indeed exhibits intrinsic quantum properties. Various theoretical proposals have been put forward to test this possibility \cite{PWZ,PWZ2,PWZ3,Anom,Anom2,Maity,Lorenci,Cho}. Most of these proposals are based on the idea that the quantum nature of gravity could be signaled by the detection of gravitons. More recently, efforts to detect gravitons have shifted from direct detection toward indirect approaches \cite{Kanno4, Anom3, Kanno3}. In such approaches, the existence of gravitons is inferred from their influence on the dynamics of a quantum system interacting with the gravitational field.

Understanding the effects of the interaction between the gravitational field and a quantum system is particularly important in the context of detectors, since operationally only the behavior of the detector itself can be directly observed. In other words, information about the gravitational field can be accessed solely through the response exhibited by the detector. As an illustration, when a detector interacts with a gravitational field, one may probe the dynamics of its quantum state or, if the detector is modeled as an oscillatory system, observe changes in its energy. Such responses potentially encode information about the nature of the gravitational field interacting with the detector. Therefore, if the gravitational field indeed possesses quantum properties, its influence is expected to manifest itself in detector responses that differ from those induced by a classical gravitational field. This detector-response–based perspective has been widely employed in various studies to relate the properties of quantized fields to observable quantities in quantum systems, including in the context of quantized gravitational fields and gravitational-wave detectors \cite{Kanno3, Sen, Marletto, Marshman}.

From a theoretical perspective, studies of quantized gravitational fields admit a variety of possible quantum states. One such state is the coherent state, which is known to exhibit dynamics that most closely resemble those of a classical field. In the context of gravitons, coherent states may arise, for example, when a linear interaction occurs between the gravitational field and matter fields \cite{Kanno8,Kanno7,Kanno9}. In addition to coherent states, another class of quantum states widely considered in studies of quantized gravity is the squeezed state. Such states possess stronger nonclassical features than coherent states. In cosmology, squeezed states naturally emerge for primordial gravitons as a consequence of inflationary dynamics, leading to the amplification of quantum fluctuations in the gravitational field \cite{Vennin, Grishchuk, Albrecht}. The existence of these different possible quantum states of the gravitational field naturally implies variations in the responses of quantum systems interacting with it. Among these states, one may therefore expect characteristic detector responses that cannot be reproduced by interactions with a purely classical gravitational field.

Based on the above discussion, in this paper we aim to investigate the response of a detector interacting with both a classical gravitational field and a quantum gravitational field. As a theoretical framework, we adopt a detector model confined in a harmonic oscillator potential, as introduced by Sen \textit{et al.} \cite{Sen}, which serves as a theoretically controlled toy model. The detector response is analyzed through the calculation of transition probabilities between its energy levels, reflecting the internal dynamics of the system induced by its interaction with the gravitational field. In this study, two distinct quantum states of the gravitational field, namely coherent states and squeezed states, are considered and compared with the classical gravitational field case.

The organization of this paper is as follows. Section \ref{sec1} provides the introduction. In Section \ref{sec2}, we describe the detector model adopted in this work, which is based on the harmonic-oscillator detector framework introduced by Sen \textit{et al.} \cite{Sen}. Section \ref{sec3} presents a general formulation of the detector transition probability, in which the gravitational-field degrees of freedom are traced out, yielding an expression that depends solely on the detector dynamics. Based on this formulation, Section \ref{sec4} computes the detector transition probability for a deterministic classical gravitational field and for a quantum gravitational field in a coherent state, and compares the resulting detector responses. Subsequently, Section \ref{sec5} discusses the detector transition probability when the gravitational field is in a squeezed state. Finally, conclusions and a summary of the main results are presented in Section \ref{sec6}. The paper is supplemented by two appendices. Appendix \ref{ApenA} contains a detailed derivation of the interaction Hamiltonian for the classical gravitational field case, while Appendix \ref{ApenB} provides an additional discussion on the properties of two-time correlations, demonstrating that the two-time correlation function of a classical gravitational field can take the same form as that of a quantum gravitational field in a coherent state.

\section{Detector Model and The Interaction Hamiltonian}\label{sec2}
In this section, we derive the form of the interaction Hamiltonian between the detector and the gravitational field treated as a quantized field. The corresponding classical-field limit will be discussed separately in Appendix \ref{ApenB}. As discussed previously, we adopt a detector model in the form of a harmonic oscillator introduced by Sen \textit{et al.} \cite{Sen}. As a first step, we consider the form of the gravitational field in the vicinity of the detector. In this study, the gravitational field is assumed to take the form of a weak gravitational wave, such that the spacetime metric can be modeled as the Minkowski metric with a small perturbation $h_{ij}$. The perturbation is assumed to satisfy the transverse–traceless gauge condition, so that the spacetime metric can be written as
\begin{eqnarray}\label{eq:2.1}
ds^2=-dt^2+\big(\delta_{ij}+h_{ij}\big)\,dx^i\,dx^j,
\end{eqnarray}
where $x^i$ are the spatial coordinates, $t$ is the time coordinate, $\delta_{ij}$ is the Kronecker delta, and $h_{ij}$ is a small perturbation that satisfies $h_{ij,i}=0$ and $h_{ii}=0$. The indices $(i,j)$ run from 1 to 3. Using this metric, the Einstein-Hilbert action up to second order can be written as follows
\begin{eqnarray}\label{eq:2.2}
S_{EH}=\frac{1}{16\,\pi\,G}\,\,\int\,d^4x\,\sqrt{-g}\,R\,\,\simeq\,\,\frac{1}{64\,\pi\,G}\,\,\int\,d^4x\,\,\Big[\dot{h}^{ij}\,\dot{h}_{ij}-h^{ij,k}\,h_{ij,k}\Big].
\end{eqnarray}
Next, by assuming that the entire system is enclosed in a box of side length $L$, the action can be expanded by discretizing the perturbation $h_{ij}$ using the following Fourier transformation
\begin{eqnarray}\label{eq:2.3}
 h_{ij}(\boldsymbol{x},t)=\frac{1}{\sqrt{\hbar G}}\sum_{\boldsymbol{k},s}\,q_{\boldsymbol{k},s}\,e^{i\boldsymbol{k}.\boldsymbol{x}}\epsilon^s_{ij}(\boldsymbol{k})
\end{eqnarray}
where $q_{\boldsymbol{k},s}$ is the mode amplitude, $\epsilon^s_{ij}(\boldsymbol{k})$ is the polarization tensor that satisfies the relation $\epsilon^A_{ij}(-\boldsymbol{k})\,\epsilon^B_{ij}(\boldsymbol{k})=\delta^{AB}$ and the index $s$ denotes linear polarization with $s=+,\times$. The normalization is chosen such that the mode variables $q_{\mathbf{k},s}$ have the dimension of length and lead to canonically normalized kinetic terms in the action. As previously explained, the entire system is assumed to be enclosed in a box of side length $L$, so the magnitude of $\boldsymbol{k}=2\pi\,n/L$ is given by $n \in \mathbb{Z}^3$. Substituting equation (\ref{eq:2.3}) into the Einstein-Hilbert action in equation (\ref{eq:2.2}), we obtain
\begin{eqnarray}\label{eq:2.4}
S_{EH}=\int\,dt\,\frac{m}{2}\,\sum_{\boldsymbol{k},s}\,\Big(\dot{q}^2_{\boldsymbol{k},s}-\boldsymbol{k}^2q^2_{\boldsymbol{k},s}\Big),
\end{eqnarray}
with $m=\frac{L^3}{16\,\pi\,\hbar\,G^2}$.

The action of a detector confined in a harmonic oscillator is modeled as an interferometer, with each mirror assumed to be a massive particle. One of the particles is chosen as the reference frame, and a coordinate system known as Fermi normal coordinates is employed, which locally represents an inertial frame. A pedagogical introduction to Fermi normal coordinates can be found in Ref. \cite{Maggiore}. The position of the second particle is defined relative to the reference particle and is expressed as $x^i(t)=\xi^i(t)$, where $\xi^i(t)$ denotes the displacement between the two particles. In Fermi normal coordinates, the spacetime metric $g_{\mu\nu}$, expanded up to second order in $x^i$, can be written as follows.
\begin{eqnarray}\label{eq:2.5}
ds^2\simeq \big(-1-R_{0i0j}\, x^i\,x^j\big)\,dt^2-\frac{4}{3}\,R_{0jik}\,x^i\,x^k\,dt\,dx^i+\bigg(\delta_{ij}-\frac{1}{3}\,R_{ikjl}\,x^k\,x^l\bigg)\,dx^i\,dx^j.
\end{eqnarray}
The Riemann tensor in this context is evaluated at the position of the particle chosen as the reference frame $\big(x^i=0\big)$. Since the detector is situated in a background consisting of gravitational waves, the Riemann tensor is computed using the Minkowski metric with a small perturbation, as given in equation (\ref{eq:2.1}). The relativistic action of the detector can then be expressed by the following equation
\begin{eqnarray}\label{eq:2.6}
S_{RD}=-m_0\,\int\,dt\,\Bigg(\sqrt{-g_{\mu\nu}\,\frac{dY^{\mu}}{dt}\frac{dY^{\nu}}{dt}}+\frac{1}{2}\,\omega^2_0\,g_{\mu\nu}\,Y^{\mu}Y^{\nu}\Bigg),
\end{eqnarray}
where $Y^{\mu}=\{t,\xi^i\}$.The second term in this action corresponds to the harmonic potential intrinsic to the detector model, with $\omega_0$ denoting the trap frequency. By substituting the metric given in equation (\ref{eq:2.5}) into the action in equation (\ref{eq:2.6}), we obtain
\begin{eqnarray}\label{eq:2.7}
S_{RD}&\simeq&\frac{m_0}{2}\,\int\,dt\,\Big(\dot{\xi^j}^2-\,R_{0i0j}(0,t)\,\xi^i\xi^j-\omega^2_0\,\delta_{jk}\,\xi^j\xi^k\Big)\nonumber\\
&=&\int\,dt\,\frac{m_0}{2}\,\Big(\,\dot{\xi^j}^2+\frac{1}{2}\,\ddot{h}_{ij}(0,t)\,\xi^i\xi^j-\omega^2_0\delta_{jk}\xi^j\xi^k\Big),
\end{eqnarray}
In the final step of the equation above, the Riemann tensor is evaluated using the metric from equation (\ref{eq:2.1}), resulting in $R_{0i0j}(0,t)=-\ddot{h}_{ij}(0,t)/2$.

The total action can be expressed as the sum of the Einstein-Hilbert action, equation (\ref{eq:2.4}), and the detector action, equation (\ref{eq:2.7}). For the detector action, the perturbation $h_{ij}$ is also expanded using the Fourier transformation given in equation (\ref{eq:2.3}). Therefore, the total action $(S=S_{EH}+S_{RD})$ becomes
 \begin{align}\label{eq:2.8}
        S=\int\,dt\,\frac{m}{2}\,\sum_{\boldsymbol{k},s}&\,\Big(\dot{q}^2_{\boldsymbol{k},s}-\boldsymbol{k}^2q^2_{\boldsymbol{k},s}\Big)\nonumber\\
        &+\int\,dt\,\frac{m_0}{2}\,\Big(\dot{\xi^j}^2+\frac{1}{\sqrt{\hbar G}}\,\sum_{\boldsymbol{k},s}\dot{q}_{\boldsymbol{k},s}\,\epsilon^s_{ij}\,\dot{\xi}^i\xi^j-\omega^2_0\,\delta_{ij}\,\xi^j\xi^k\Big).
 \end{align}
 As in the study by Sen \textit{et al.} \cite{Sen},  the interferometer considered in this work is modeled as a one-dimensional system.  This reduction corresponds to a physical situation in which one arm of the interferometer  dominates the dynamics, while the remaining transverse degrees of freedom are assumed to be frozen. As a consequence, the oscillatory motion of the detector, confined by the harmonic potential,  is effectively one-dimensional. In addition, the gravitational wave is assumed to propagate along a single direction,  taken to be the $z$-direction, with $|\boldsymbol{k}|=\omega$,  and to possess only a single polarization mode, namely the $+$ polarization. Under these assumptions, both the detector motion and the gravitational-wave coupling  effectively reduce to a single dynamical mode,  so that the mode variable $q_{\boldsymbol{k},s}$ can be simplified as 
$\mathfrak{R}(q_{\boldsymbol{k},z},+)=q$. By further neglecting fluctuations along the $y$- and $z$-directions, 
the action in equation (\ref{eq:2.8}) can be simplified to
 \begin{eqnarray}
        S=\int\,dt\bigg(\frac{m}{2}\big(\dot{q}^2-\omega^2q^2\big)+\frac{m_0}{2}\Big(\dot{\xi}^2-\frac{2\mathcal{G}\dot{q}\dot{\xi}\xi}{m_0}-\omega^2_0\xi^2\Big)\bigg)
 \end{eqnarray}
Where $\mathcal{G}=\frac{m_0}{2\sqrt{\hbar G}}$ and $\xi^x=\xi$. Based on this action, the corresponding Hamiltonian can be written as
 \begin{eqnarray}\label{eq:2.10}
 H=\frac{\frac{p^2}{2m}+\frac{\pi^2}{2m_0}+\frac{\mathcal{G}p\pi\xi}{m\,m_0}}{1-\frac{\mathcal{G}^2\xi^2}{m\,m_0}}+\frac{1}{2}\,m\,\omega^2\,q^2+\frac{1}{2}\,m_0\,\omega^2_0\,\xi^2.
 \end{eqnarray}
 with
  \begin{eqnarray}
        p=\frac{\partial L}{\partial \dot{q}}=m\dot{q}-\mathcal{G}\dot{\xi}\xi\,\,\,\,\,\,\,\,\,\,\,\,\,\,\,\,\,\,\,\,\,\,\,\,\pi=\frac{\partial L}{\partial \dot{\xi}}=m_0\dot{\xi}-\mathcal{G}\dot{q}\xi.
  \end{eqnarray}
  
Throughout this work we restrict ourselves to the weak-coupling regime, where terms of order $\mathcal{O}(\mathcal{G}^2)$ and higher can be consistently neglected. The Hamiltonian in equation (\ref{eq:2.10}) can be expanded and truncated at order $\mathcal{O}(\mathcal{G})$, yielding
\begin{eqnarray}\label{eq:2.12}
        H\simeq\frac{p^2}{2m}+\frac{1}{2}\,m\omega^2q^2+\frac{\pi^2}{2m_0}+\frac{1}{2}\,m_0\omega_0^2\xi^2+\frac{\mathcal{G}}{m\,m_0}\,p\pi\xi.
\end{eqnarray}
This Hamiltonian describes two harmonic oscillators, corresponding respectively to the oscillatory degrees of freedom of the detector and the gravitational wave, together with an interaction term that encodes their mutual coupling. The interaction is assumed to take place over a finite time interval from $t_i$ to $t_f$, during which the coupling constant $\mathcal{G}$ is switched on and off adiabatically. The Hamiltonian is then quantized by promoting $p, q, \xi$, and $\pi$ to operators satisfying the canonical commutation relations $[\hat{\xi},\hat{\pi}]=[\hat{q},\hat{p}]=i\hbar$. In terms of annihilation and creation operators, these operators can be written as follows.
\begin{eqnarray}
        \hat{q}&=&\sqrt{\frac{\hbar}{2m\omega}}\Big(\hat{a}+\hat{a}^{\dagger}\Big)\,\,\,\,\,\,\,\,\,\,\,\,\,\,\,\,\,\,\,\,\,\,\,\hat{p}=i\sqrt{\frac{m\hbar\omega}{2}}\Big(\hat{a}^{\dagger}-\hat{a}\Big)\\
        \nonumber\\
        \hat{\xi}&=&\sqrt{\frac{\hbar}{2m_0\omega_0}}\Big(\hat{\chi}+\hat{\chi}^{\dagger}\Big)\,\,\,\,\,\,\,\,\,\,\,\,\,\,\,\,\,\hat{\pi}=i\sqrt{\frac{m_0\hbar\omega_0}{2}}\Big(\hat{\chi}^{\dagger}-\hat{\chi}\Big)\label{eq:2.14}
\end{eqnarray}
Where $[\hat{a},\,\hat{a}^{\dagger}]=[\hat{\chi},\,\hat{\chi}^{\dagger}]=1$. Then, based on these annihilation and creation operators, the eigenstates of the number operators $\hat{n}_{G}=\hat{a}^{\dagger}\,\hat{a}$ and $\hat{n}_r=\hat{\chi}^{\dagger}\,\hat{\chi}$ can be defined as
\begin{eqnarray}
\hat{n}_G\,\ket{n_G}=n_G\,\ket{n_G},\,\,\,\,\,\,\,\,\,\,\,\,\,\,\,\,\,\,\,\,\,\,\,\,\,\,\,\,\,\,\,\,\,\,\,\,\hat{n}_r\,\ket{n_r}=n_r\,\ket{n_r}.
\end{eqnarray}
The Hamiltonian in equation (\ref{eq:2.12}) can be separated into two parts, $\hat{H}=\hat{H}_0+\hat{H}^{\text{int}}$. In the operator representation, the interaction Hamiltonian $\hat{H}^{\text{int}}$ can be written as follows
\begin{eqnarray}\label{eq:2.16}
        H^{\text{int}}=\frac{\mathcal{G}}{2m\,m_0}\hat{p}\otimes\Big(\hat{\xi}\hat{\pi}+\hat{\pi}\hat{\xi}\Big).
\end{eqnarray}
From this result, it follows that the gravitational field couples to the detector through its conjugate momentum $\hat{p}$, a feature that will play a central role in determining the structure of the detector transition probabilities discussed in the subsequent sections. Using this interaction Hamiltonian, the transition probability of the detector for the case of a quantized gravitational field will be calculated.

\section{Detector Transition Probability: General Framework}\label{sec3}
In this section, we discuss the general form of the detector transition probability after tracing out the degrees of freedom associated with the gravitational field (or gravitons) and establish a general framework that will be used throughout this work. As a first step, we review the general expression for the transition probability of the combined detector–graviton system, which can be written as
\begin{equation}\label{eq:3.1}
P_{i\to f}=\Big|\bra{\psi_f}\hat{U}_I(t)\ket{\psi_i}\Big|^2.
\end{equation}
Here, $\ket{\psi_i}$ and $\ket{\psi_f}$ represent the initial and final quantum states of the combined system consisting of the detector and the gravitational field, respectively. These states belong to the total Hilbert space $\mathcal{H}_G \otimes \mathcal{H}_r$, where $\mathcal{H}_G$ denotes the Hilbert space of the gravitational field and $\mathcal{H}_r$ denotes the Hilbert space of the detector. Meanwhile, $\hat U_I(t)$ is the unitary time-evolution operator generated by the interaction Hamiltonian in equation (\ref{eq:2.16}) in the interaction picture.

The initial quantum state of the detector is assumed to be in the energy eigenstate labeled by $n_r$, denoted by $\ket{n_r}$. Meanwhile, the initial state of the gravitational field (gravitons) is taken to be in a general form $\ket{\psi_{\rm GW}}$. In order for the transition probability in equation (\ref{eq:3.1}) to represent solely the detector dynamics, the gravitational-field degrees of freedom in the final state are summed over all available states. Accordingly, the final state of the gravitational field is expressed as a sum over a complete basis $(\sum_{m_G}\,\ket{m_G})$. If the final state of the detector is $\ket{n_r'}$ (where $n'_r\neq n_r$), the detector transition probability can then be written as
\begin{eqnarray}\label{eq:3.2}
P_{i\to f}&=&\sum_{m_G}\,\Big(\bra{m_G}\otimes\bra{n'_r}\Big)\,\hat{U}_I(t)\,\Big(\ket{\psi_{\rm GW}}\otimes\ket{n_r}\Big)\,\Big(\bra{\psi_{\rm GW}}\otimes\bra{n_r}\Big)  \,\hat{U}^{\dagger}_I(t)\,\Big(\ket{m_G}\otimes\ket{n'_r}\Big)\nonumber\\
&=&tr_G\Big\{ \bra{n'_r}\,\hat{U}_I(t)\,\Big(\ket{\psi_{\rm GW}}\bra{\psi_{\rm GW}}\otimes\ket{n_r}\bra{n_r}\Big)\,\hat{U}^{\dagger}_I(t)\,\ket{n'_r} \Big\}.
\end{eqnarray}
Since the gravitational-field degrees of freedom are traced out in this expression, the resulting transition probability fully characterizes the dynamics, or response, of the detector when interacting with a quantized gravitational field.

Next, the unitary time-evolution operator $\hat U_I(t)$ is expanded to first order in the interaction Hamiltonian $\hat H_I^{\rm int}(t)$, which is sufficient in the weak-coupling regime considered here, as
\begin{eqnarray}\label{eq:3.3}
\hat{U}_I(t)&=&\mathcal{T}\,\exp\Big\{ -\frac{i}{\hbar}\,\int^t_0\,dt'\,\hat{H}^{\text{int}}_I(t')\Big\}\nonumber\\
&\simeq&1-\frac{i}{\hbar}\int^t_0\,dt' \hat{H}^{\text{int}}_I(t').
\end{eqnarray}
Here, $\hat H_I^{\rm int}(t)$ denotes the interaction Hamiltonian in equation (\ref{eq:2.16}) expressed in the interaction picture, which is given by
\begin{eqnarray}\label{eq:3.6}
\hat{H}^{\text{int}}_I(t)&=&e^{\frac{i}{\hbar}\hat{H}_0\,t}\,\hat{H}^{\text{int}}\,e^{-\frac{i}{\hbar}\hat{H}_0\,t}\nonumber\\
&=&\frac{\mathcal{G}}{2m\,m_0}\,\,\hat{p}_I(t)\otimes\Big(\hat{\xi}_I(t)\,\hat{\pi}_I(t)+\hat{\pi}_I(t)\,\hat{\xi}_I(t)\Big),
\end{eqnarray}
where $\hat p_I(t), \hat \xi_I(t)$, and $\hat \pi_I(t)$ denote the operators $\hat{p}, \hat{\xi}$, and $\hat{\pi}$ in the interaction picture, respectively. Substituting equation (\ref{eq:3.3}) into equation (\ref{eq:3.2}), the transition probability becomes
\begin{eqnarray}
P_{i\to f}&=&\frac{\mathcal{G}^2}{4\hbar^2 m^2 m^2_0}\,\int^t_0dt_1\,\int^t_0dt_2\,\,\bra{\psi_{GW}}\hat{p}_I(t_1)\,\hat{p}_I(t_2)\ket{\psi_{GW}}\nonumber\\
&\,\,&\otimes\bra{n_r}\big(\hat{\xi}_I(t_1)\hat{\pi}_I(t_1)+\hat{\pi}_I(t_1)\hat{\xi}_I(t_1)\big)\ket{n'_r}\bra{n'_r}\big(\hat{\xi}_I(t_2)\hat{\pi}_I(t_2)+\hat{\pi}_I(t_2)\hat{\xi}_I(t_2)\big)\ket{n_r}.
\end{eqnarray}
Using equation (\ref{eq:2.14}), the detector matrix element can be written as
\begin{align}
\bra{n_r}\big(\hat{\xi}_I(t_1)&\hat{\pi}_I(t_1)+\hat{\pi}_I(t_1)\hat{\xi}_I(t_1)\big)\ket{n'_r}\nonumber\\
&=-i\hbar\,\Big(\sqrt{(n_r+1)(n_r+2)}\,\delta_{n'_r,\,n_r+2}-\sqrt{n_r(n_r-1)}\,\delta_{n'_r,\,n_r-2}\Big)\,e^{-i\,(n'_r-n_r)\,\omega_0\,t_1}.
\end{align}
Accordingly, the transition probability can be expressed as
\begin{eqnarray}\label{eq:3.7}
P_{i\to f}&=&\frac{\mathcal{G}^2}{4m^2 m^2_0}\,\,\Big|\sqrt{(n_r+1)(n_r+2)}\,\delta_{n'_r,\,n_r+2}-\sqrt{n_r(n_r-1)}\,\delta_{n'_r,\,n_r-2}\Big|^2\nonumber\\
&\,\,&\times \int^t_0dt_1\,\int^t_0dt_2\,\,\langle\hat{p}_I(t_1)\,\hat{p}_I(t_2)\rangle\,e^{i\,(n'_r-n_r)\,\omega_0\,(t_2-t_1)}.
\end{eqnarray}

From this result, it is evident that the influence of the gravitational field on the detector transition probability is entirely encoded in the two-time correlation function of the conjugate momentum, namely $\langle \hat{p}_I(t_1)\hat{p}_I(t_2)\rangle$. This correlation function encapsulates all the information about the quantum state of the gravitational field that is relevant for the detector dynamics. Using this expression, the detector transition probability can be evaluated for different choices of gravitational-field states. In the following section, we first analyze the case in which the gravitational field is treated as a classical field. The result is then compared with the case of a quantum gravitational field in a coherent state, which represents the quantum limit closest to a classical field description.

\section{Classical and Coherent Gravitational Fields}\label{sec4}
In this section, we compute the detector transition probability by treating the gravitational field as a quantum field in a coherent state.The discussion is organized into three stages. First, we evaluate the detector transition probability for the case of a deterministic classical gravitational field. Next, we analyze the transition probability when the gravitational field is in a coherent state. Finally, the results of these two calculations are compared in order to highlight the correspondences and differences that arise between them.

\subsection{Classical gravitational field}\label{subsec4.1}
To compute the detector transition probability when the gravitational field is treated as a classical field, the interaction Hamiltonian in equation (\ref{eq:2.16}) cannot be used directly, since in that formulation the gravitational field is promoted to a quantum operator. In the classical description, the gravitational field is instead treated as a prescribed time-dependent function. We therefore employ an interaction Hamiltonian that retains the same coupling structure as its quantum counterpart, while keeping the gravitational field as a classical quantity rather than promoting it to an operator. Mathematically, the interaction Hamiltonian in the interaction picture can be written as
\begin{eqnarray}
\hat{H}^{\text{int}}_I(t)=\frac{\dot{h}_{+}(t)}{4}\,\Big(\hat{\pi}_I(t)\,\hat{\xi}_I(t)+\hat{\xi}_I(t)\,\hat{\pi}_I(t)\Big).
\end{eqnarray}
A detailed derivation of this interaction Hamiltonian is provided in Appendix \ref{ApenA}.

The detector transition probability can then be computed by following a procedure analogous to that leading to equation (\ref{eq:3.7}). The difference in the present case lies in the fact that the field coupled to the detector is a classical field, $\dot{h}_{+}(t)$. Accordingly, the relevant two-time correlation function is given by $\langle \dot{h}_{+}(t_1)\dot{h}_{+}(t_2)\rangle$. The detector transition probability can therefore be written as
\begin{eqnarray}\label{eq:4.2}
P_{i\to f}&=&\frac{1}{16}\,\,\Big|\sqrt{(n_r+1)(n_r+2)}\,\delta_{n'_r,\,n_r+2}-\sqrt{n_r(n_r-1)}\,\delta_{n'_r,\,n_r-2}\Big|^2\nonumber\\
&\,\,&\times \int^t_0dt_1\,\int^t_0dt_2\,\,\langle\dot{h}_{+}(t_1)\,\dot{h}_{+}(t_2)\,\rangle\,e^{i\,(n'_r-n_r)\,\omega_0\,(t_2-t_1)}.
\end{eqnarray}
In this section, the classical field $h_{+}(t)$ is assumed to be a deterministic classical field modeled as a one-dimensional gravitational wave that is periodic, linear, and contains only the plus polarization. Explicitly, the field is taken to be $h_{+}(t)=2 f_0 \cos(\omega t)$, where $f_0$ denotes the gravitational-wave amplitude. Since the field $h_{+}(t)$ is deterministic, the two-time correlation function reduces to an ordinary product, $\langle \dot{h}_{+}(t_1)\dot{h}_{+}(t_2) \rangle= \dot h_{+}(t_1)\dot h_{+}(t_2)$, which in the classical case is not a quantum expectation value but simply the product of classical field amplitudes. As a result, the transition probability in equation (\ref{eq:4.2}) can be expressed as
\begin{eqnarray}\label{eq:4.3}
P_{i\to f}&=&\frac{f^2_0\,\omega^2}{4}\,\,\Big|\sqrt{(n_r+1)(n_r+2)}\,\delta_{n'_r,\,n_r+2}-\sqrt{n_r(n_r-1)}\,\delta_{n'_r,\,n_r-2}\Big|^2\nonumber\\
&\,\,&\times\left\{\frac{\sin^2\left[\,\left((n'_r-n_r)\,\omega_0-\omega\right)\,\frac{t}{2}\right]}{\left((n'_r-n_r)\,\omega_0-\omega\right)^2}+\frac{\sin^2\left[\left((n'_r-n_r)\,\omega_0+\omega\right)\,\frac{t}{2}\right]}{\left((n'_r-n_r)\,\omega_0+\omega\right)^2}\right\}.
\end{eqnarray}
In performing the integrations, the cross terms arising from the product $\dot{h}_{+}(t_1)\dot{h}_{+}(t_2)$ can be neglected. A detailed justification for neglecting these terms is provided in Appendix \ref{ApenC}. These terms contain rapidly oscillating factors whose double-time integrals remain bounded and do not grow with the interaction time, and therefore they become negligible compared to the resonant contributions in the long interaction-time limit. Assuming that the interaction between the gravitational wave and the detector persists for a sufficiently long duration, the transition probability in equation (\ref{eq:4.3}) reduces to
\begin{eqnarray}\label{eq:4.4}
P_{i\to f}&=&\frac{f^2_0\,\omega^2\,\pi\,t}{8}\,\,\Big|\sqrt{(n_r+1)(n_r+2)}\,\delta_{n'_r,\,n_r+2}-\sqrt{n_r(n_r-1)}\,\delta_{n'_r,\,n_r-2}\Big|^2\nonumber\\
&\,\,&\times\Big\{\delta\left((n'_r-n_r)\,\omega_0-\omega\right)+\delta\left((n'_r-n_r)\,\omega_0+\omega\right)\Big\}.
\end{eqnarray}

In this expression for the transition probability, two distinct Dirac delta functions appear. These two delta functions represent two resonance conditions under which the transition probability attains its maximum value. The first resonance condition occurs when the energy level of the detector in the final state is lower than that in the initial state by an amount $\hbar\omega$, namely when $E_{n_r'} = E_{n_r} - \hbar\omega$. In this case, the interaction between the detector and the classical gravitational field leads to a decrease in the detector’s energy that is equal to the frequency of the gravitational wave. As an illustration, if the detector is initially in the energy level $n_r = 2$ and undergoes a transition to the final state $n_r' = 0$, equation (\ref{eq:4.4}) reduces to
\begin{eqnarray}\label{eq:4.5}
P_{2\to0}=\frac{f^2_0\,\omega^2\,\pi\,t}{4}\,\,\delta\big(\omega-2\omega_0\big).
\end{eqnarray}
The second resonance condition arises when the energy level of the detector in the final state is higher than that in the initial state by $\hbar\omega$, i.e., when $E_{n_r'} = E_{n_r} + \hbar\omega$. In this situation, the detector gains energy as a result of its interaction with the classical gravitational wave. For example, for a transition from the initial state $n_r = 0$ to the final state $n_r' = 2$, the transition probability is given by
\begin{eqnarray}\label{eq:4.6}
P_{0\to2}=\frac{f^2_0\,\omega^2\,\pi\,t}{4}\,\,\delta\big(2\omega_0-\omega\big).
\end{eqnarray}
These two results will serve as classical benchmarks and will subsequently be compared with the detector transition probabilities obtained when the gravitational field is treated as a quantum field in a coherent state. Before carrying out this comparison, we first compute the transition probability for the coherent-state case.

\subsection{Coherent-state gravitational field}\label{subsec4.2}
In this section, we compute the detector transition probability by assuming that the gravitational field is treated as a quantum field in a coherent state. A coherent state is a quantum state whose expectation-value dynamics most closely resemble those of a classical field. Mathematically, a coherent state is defined as an eigenstate of the annihilation operator, satisfying the relation $\hat a \ket{\alpha} = \alpha \ket{\alpha}$, where $\alpha$ is a complex number. A coherent state can be generated by applying the displacement operator to the vacuum state, namely
 \begin{eqnarray}
        \ket{\alpha}&=&\hat{D}(\alpha)\ket{0}\nonumber\\
        &=&e^{-\frac{|\alpha|^2}{2}}\,\sum^{\infty}_{n_G=0}\,\frac{\alpha^{n_G}}{\sqrt{n_G!}}\,\ket{n_G}.
 \end{eqnarray}
The displacement operator $\hat D(\alpha)$ is defined as
\begin{eqnarray}
\hat{D}(\alpha)=\exp\Big[\alpha\,\hat{a}^{\dagger}-\alpha^*\,\hat{a}\Big].
\end{eqnarray}
The parameter $\alpha$ can be written in polar form as $\alpha = |\alpha| e^{i\varphi}$. Physically, $|\alpha|$ determines the magnitude of the displacement of the vacuum state in phase space, while $\varphi$ represents the associated phase angle. A coherent state has an average field-quanta (graviton) number given by $\braket{n_G} = \bra{\alpha}\hat{n}_G\ket{\alpha} = |\alpha|^2$. General discussions of the properties of coherent states can be found, for example, in Refs. \cite{Knight,Fox}. In the context of quantum gravity, graviton coherent states may arise due to linear couplings between metric perturbations and matter fields. One such example is provided by gravitons produced in the early universe \cite{Kanno8,Kanno7}. In addition, graviton coherent states have also been reported to emerge in black-hole systems \cite{Kanno9, Manikandan}.

The detector transition probability for a quantized gravitational field can be computed using equation (\ref{eq:3.7}). In that expression, the quantum state of the gravitational field enters the calculation through the two-time correlation function $\langle \hat p_I(t_1)\hat p_I(t_2)\rangle=\bra{\psi_{\rm GW}} \hat p_I(t_1)\hat p_I(t_2) \ket{\psi_{\rm GW}}$. In this case, the gravitational field is assumed to be in a coherent state, so that $\ket{\psi_{\rm GW}}\equiv\ket{\alpha}$. By exploiting the properties of the displacement operator, namely $\hat{D}^{\dagger}(\alpha)\hat{a}\hat{D}(\alpha)=\hat{a}+\alpha$ and $\hat{D}^{\dagger}(\alpha)\hat{a}^{\dagger}\hat{D}(\alpha)=\hat{a}^{\dagger}+\alpha^*$, the two-time correlation function can be written as 
\begin{eqnarray}
\langle\hat{p}_I(t_1)\,\hat{p}_I(t_2)\rangle&=&\bra{0}\hat{D}^{\dagger}(\alpha)\,\hat{p}_I(t_1)\,\hat{p}_I(t_2)\,\hat{D}(\alpha)\ket{0}\nonumber\\
&=&\frac{m\hbar\omega}{2}\,\bigg\{(\alpha^*\,e^{i\omega t_1}-\alpha\,e^{-i\omega\,t_1})(\alpha e^{-i\omega t_2}-\alpha^*\,e^{i\omega t_2})+e^{i\omega(t_2-t_1)}\bigg\}.
\end{eqnarray}
Substituting this result into equation (\ref{eq:3.7}), and assuming that the interaction time between the detector and the gravitational wave is sufficiently long, the detector transition probability can be expressed as
\begin{eqnarray}\label{eq:4.10}
P_{i\to f}&=&\frac{G^2\,\pi\,\hbar\,\omega t}{4m\,m^2_0}\,\,\Big|\sqrt{(n_r+1)(n_r+2)}\,\delta_{n'_r,\,n_r+2}-\sqrt{n_r(n_r-1)}\,\delta_{n'_r,\,n_r-2}\Big|^2\nonumber\\
&\,\,&\times\Big\{ |\alpha|^2\,\delta\left((n'_r-n_r)\,\omega_0-\omega\right)+(|\alpha|^2+1)\,\delta\left((n'_r-n_r)\,\omega_0+\omega\right)\Big\}.
\end{eqnarray}
As in the classical case, the cross terms arising from $(\alpha^* e^{i\omega t_1}-\alpha e^{-i\omega t_1})(\alpha e^{-i\omega t_2}-\alpha^* e^{i\omega t_2})$ are neglected. These terms contain rapidly oscillating factors whose double-time integrals remain bounded and do not grow with the interaction time, in contrast to the resonant contributions which increase with time.

Equation (\ref{eq:4.10}) shows that the detector transition probability for a quantum gravitational field in a coherent state also exhibits two resonance conditions at which the transition probability attains its maximum value. However, unlike the classical description, in the present case the gravitational field is promoted to a quantum field. Consequently, when the detector undergoes a decrease in energy by two energy levels, the released energy can be associated with an excitation of the quantized gravitational field. Physically, this process can be interpreted as the emission of a single quantum of the gravitational field (a graviton). As an illustration, this situation occurs for a transition from the initial state $n_r = 2$ to the final state $n_r' = 0$, for which the transition probability is given by
\begin{eqnarray}\label{eq:4.11}
P_{2\to 0}&=&\frac{G^2\,\pi\,\hbar\,\omega t}{2m\,m^2_0}\,(|\alpha|^2+1)\,\delta\left(\omega-2\omega_0\right).
\end{eqnarray}
Conversely, when the detector undergoes an increase in energy by two levels, the process can occur due to the absorption of energy from the quantized gravitational field. Within the framework of a quantized-field description, this process may be interpreted as the annihilation of a single quantum of the gravitational field (a graviton), whose energy is absorbed by the detector. As an example, this situation arises when the detector is initially in the state $n_r = 0$ and transitions to the final state $n_r' = 2$. The corresponding transition probability is given by
\begin{eqnarray}\label{eq:4.12}
P_{0\to 2}&=&\frac{G^2\,\pi\,\hbar\,\omega t}{2m\,m^2_0}\,|\alpha|^2\,\delta\left(2\omega_0-\omega\right).
\end{eqnarray}
In the following, the transition probabilities in equations (\ref{eq:4.11}) and (\ref{eq:4.12}) will be compared with the detector transition probabilities previously obtained for the classical gravitational field case.

\subsection{Comparison With The Classical Field}\label{subsec4.3}
Based on the previous calculations, for both the classical gravitational field and the quantum gravitational field in a coherent state, the detector transition probabilities exhibit two resonance conditions, namely when the transition occurs from  $n_r = 0$ to $n_r' = 2$ and from $n_r = 2$ to $n_r' = 0$. These two resonance conditions are compared between the cases of a classical gravitational field and a coherent gravitational field. To enable this comparison, an explicit relation between the amplitude of the classical gravitational field $f_0$ and the parameters characterizing the quantum gravitational field is required. In this work, such a relation is derived by following the procedure introduced by Sen \textit{et al.} \cite{Sen}. As a first step, we consider the energy carried by the gravitational wave interacting with the detector. Within this framework, the gravitational wave energy available for interaction with the detector is associated with the energy flux of the gravitational wave passing through the effective area of the detector. In particular, the total energy of the gravitational wave passing through an area element $dA$ can be written as
\cite{Maggiore}
 \begin{eqnarray}
        \frac{d\,E}{d\,A}=\frac{1}{32\pi G}\int^{\infty}_{0}\,dt\,\braket{\dot{h}_{ij}\,\dot{h}_{ij}}.
 \end{eqnarray}
The above expression assumes that the entire system is enclosed within a sphere of radius $R$, where 
$dA = R^2\,d\Omega$, representing the energy flowing into or out of the system through a solid angle $d\Omega$. To carry out the integration, a periodic plane-polarized gravitational wave is considered, given by 
$h_{ij}(t) = h_+(t) = 2f_0 \cos{\omega t}$. Thus, the time duration can be simplified to a single oscillation period, from 
$t_0$ to $t = 2\pi/\omega$. Hence, the integration can be performed straightforwardly, as the temporal average corresponds to an average over a constant parameter, yielding
 \begin{eqnarray}\label{eq:4.14}
 E&=&\frac{r^2}{8G}\int^{2\pi/\omega}_0 dt\,\braket{\dot{h}^2_{+}(t)}\nonumber\\
 &=&\frac{\pi\omega\,f^2_{\text{0}}\,r^2}{2G}\nonumber\\
 &=&\frac{\pi\omega\,f^2_{\text{0}}\,L^2}{4G}
 \end{eqnarray}
where $L$ is the length of a box assumed to be inside a sphere of radius $R$, such that $L = R/\sqrt{2}$. 

Unlike the study by S. Sen et al. \cite{Sen}, where the energy of the quantized gravitational wave is interpreted as the energy of $n_G$ gravitons with frequency $\omega_0$, corresponding to an initial number state $\ket{n_G}$, the present study considers gravitons in a coherent state $\ket{\alpha}$. Consequently, the energy is taken as the average energy of a coherent state, given by $E = \braket{n_G}\hbar\omega = |\alpha|^2\hbar\omega$. Therefore, the expression for the squared amplitude $f_0^2$ can be written as follows
\begin{eqnarray}\label{eq:4.15}
        f^2_{0}&=&\frac{4\,E\,G}{\pi\,\omega\,L^2}\nonumber\\
        &=&\frac{4\,\hbar\,G}{\pi\,L^2}\,|\alpha|^2.
\end{eqnarray}
From this result, it is evident that the amplitude of the classical gravitational field $f_0$ is proportional to the magnitude of the coherent-state displacement parameter, $|\alpha|$. This relation provides a clear identification between the amplitude of the classical gravitational field and the parameter characterizing the coherent state
of the quantized gravitational field. Using this definition of the amplitude $f_0$, the detector transition probabilities for a quantum gravitational field in a coherent state, given by equations (\ref{eq:4.11}) and (\ref{eq:4.12}), can be directly compared with the detector transition probabilities obtained for a classical gravitational field, as expressed in
equations (\ref{eq:4.5}) and (\ref{eq:4.6}).

The transition probability in equation (\ref{eq:4.12}), when expressed in terms of the classical field amplitude $f_0$, can be written as
\begin{eqnarray}\label{eq:4.16}
P_{0\to 2}&=&\frac{f^2_0\,\pi^3\,\omega\,t}{2\,L}\,\,\delta\big(2\omega_0-\omega\big)\nonumber\\
&=&\frac{f^2_0\,\pi\,\omega^2\,t}{4}\,\,\delta\big(2\omega_0-\omega\big)
\end{eqnarray}
In deriving the above expression, it is assumed that the box length $L$ is sufficiently large so that it can be associated with half the circumference of a sphere enclosing the propagation volume of the gravitational wave. The radius of this sphere is taken to be equal to the distance traveled by the gravitational wave during one oscillation period, $t = 2\pi/\omega$. Under this assumption, one obtains the approximation $L \simeq 2\pi^{2}/\omega$. The form of the transition probability in equation (\ref{eq:4.16}) is identical to the result given in equation (\ref{eq:4.6}). This indicates that, for the process in which the detector undergoes an upward transition in its energy levels, the detector response induced by a quantum gravitational field in a coherent state is indistinguishable from that induced by a deterministic classical gravitational field.

The transition probability in equation (\ref{eq:4.11}), when expressed in terms of the classical field amplitude $f_0$ using the same procedure, can be written as
\begin{eqnarray}\label{eq:4.17}
P_{2 \to 0}&=&\frac{f^2_0\,\pi\,\omega^2\,t}{4}\,\,\delta \big(\omega-2\omega_0\big)+\frac{2\pi^2\,\hbar\,\omega\,t\,G}{L^3}\,\,\delta\big(\omega-2\omega_0\big)\nonumber\\
&=&\frac{f^2_0\,\pi\,\omega^2\,t}{4}\,\,\delta \big(\omega-2\omega_0\big)+\frac{\hbar\,\omega^4\,t\,G}{4\,\pi^4}\,\,\delta\big(\omega-2\omega_0\big).
\end{eqnarray}
When compared with the detector transition probability given in equation (\ref{eq:4.5}), it is evident that the transition probability in equation (\ref{eq:4.17}) contains an additional term that is absent in the case of a deterministic classical gravitational field. This indicates that, for the process in which the detector undergoes a downward transition in its energy levels, the detector response induced by a quantum gravitational field in a coherent state differs from that induced by a deterministic classical gravitational field. This distinction becomes transparent when the two-time correlation function of the coherent state is decomposed explicitly. Nevertheless, the presence of this additional term is not sufficient to be uniquely associated with quantum gravitational effects. This is because the additional contribution can, in principle, also be reproduced by a classical gravitational field, provided that the field possesses an appropriate two-time correlation structure. To clarify this point, we explicitly show below how the additional term in equation (\ref{eq:4.17}) arises from the two-time correlation function of a coherent state. In general, the two-time correlation function of the gravitational field in a coherent state can be written as
\begin{equation}
\begin{aligned}
\bra{\alpha}\,\hat{p}_I(t_1)\,\hat{p}_I(t_2)\,\ket{\alpha}
&=
\underbrace{
\bra{\alpha} \hat{p}_I(t_1)\ket{\alpha}\,\bra{\alpha} \hat{p}_I(t_2)\,\ket{\alpha}
}_{\text{mean-field}}\,\,
+
\underbrace{
F^{(\rm coh)}(t_2-t_1)
}_{\text{difference-time correlation function}},
\end{aligned}
\end{equation}
Here, the mean-field term gives rise to the first contribution in the transition probability in equation (\ref{eq:4.17}), which directly corresponds to the response obtained for a deterministic classical gravitational field. The function $F^{(\mathrm{coh})}(t_2-t_1)$, on the other hand, represents a correlation that depends on the time difference, which in the present case takes the form $F^{(\mathrm{coh})}(t_2-t_1)=e^{i\omega(t_2-t_1)}$. This term is responsible for the additional contribution appearing in equation (\ref{eq:4.17}). For comparison, in the case of a classical gravitational field defined as a purely deterministic field, the two-time correlation function contains only
the mean-field contribution. However, if the classical gravitational field is extended by including a fluctuating component on top of its deterministic part, a two-time correlation function with a structure similar to $F^{(\mathrm{coh})}(t_2-t_1)$ can also be generated. An explicit example of this situation is presented in
Appendix \ref{ApenB}. Therefore, the transition probability in equation (\ref{eq:4.17}) can, in principle, also be reproduced within a classical gravitational field framework whose correlations depend only on the time difference (i.e., stationary classical fields).

In the following section, we turn to quantum states of the gravitational field whose two-time correlation functions contain contributions beyond the mean-field term and functions that depend only on the time difference, and therefore cannot be reproduced by the two-time correlation functions of stationary classical gravitational fields. One such quantum state that satisfies this criterion is the squeezed state. We then explicitly compute how these non-classical correlations contribute to the detector transition probability and demonstrate how their effects differ in principle from detector responses that can be generated within a stationary classical gravitational-field framework.

\section{Non-classical Gravitational Fields: Squeezed States}\label{sec5}
Beyond coherent states, a quantized gravitational field may occupy various other quantum states. One particularly well-motivated class is the squeezed state, which can naturally arise when gravitons originate from the early Universe  \cite{Vennin, Grishchuk, Albrecht}. As a consequence of cosmological inflation, the initial quantum fluctuations of the gravitational field undergo parametric amplification, causing the graviton quantum state to evolve into a squeezed state. Gravitons produced through this mechanism are commonly referred to as primordial gravitons. In a cosmological context, primordial gravitons typically appear in the form of two-mode squeezed states. However, in the present work we restrict our analysis to the case of a single-mode squeezed state. This restriction is directly related to the assumptions of the model employed, in which the gravitational wave in the vicinity of the detector is assumed to propagate along a single direction (one-dimensional propagation). Under this assumption, the gravitational-field degrees of freedom that couple to the detector can be reduced to an effective single mode. This simplification does not remove the physical content of the results, since the essential non-classical correlations inherent to squeezed states remain preserved within the single-mode description. Mathematically, the single-mode squeezed state can be written as follows
\begin{eqnarray}
        \ket{\zeta}&=&\hat{S}(\zeta)\ket{0}\nonumber\\
        &=&\frac{1}{\sqrt{\cosh r}}\,\sum^{\infty}_{n_G=0}\,(-1)^{n_G}\,\frac{\sqrt{(2n_G)!}}{2^{n_G}\,n_G!}\,e^{in_G\theta}\,(\tanh r)^{n_G}\,\ket{2n_G}.
\end{eqnarray}
Here, $\hat{S}(\zeta)$ denotes the single-mode squeezing operator, which is mathematically defined as
\begin{eqnarray}
        \hat{S}(\zeta)=\exp\bigg[\frac{1}{2}\,\,\big(\zeta^*\,\hat{a}^2-\zeta\,\hat{a}^{\dagger\,2}\big)\bigg],\,\,\,\,\,\,\,\,\,\,\,\,\,\,\,\,\,\,\,\,\,\,\,\,\,\,\,\zeta=r\,e^{i\theta},
\end{eqnarray}
The parameters $r$ and $\theta$ represent the squeezing amplitude and the squeezing phase, respectively. The parameter $r$ quantifies the strength of the squeezing experienced by the field, while the phase $\theta$ determines the squeezing direction in phase space. In the present analysis, the squeezing parameter is taken in the range $0 \leq r < \infty$, whereas the squeezing phase $\theta$ is assumed to lie within the interval $0 \leq \theta < 2\pi$. General discussions of squeezed states can be found in Refs. \cite{Knight,Fox}.

Next, we compute the two-time correlation function of the field $\hat{p}_I(t)$ by assuming that the quantum state of the gravitational field is a squeezed state, $\ket{\psi_{\rm GW}} \equiv \ket{\zeta}$. Using the transformation properties of the squeezing operator, $\hat{S}^{\dagger}(\zeta)\,\hat{a}\,\hat{S}(\zeta)=\hat{a}\,\cosh r-\hat{a}^{\dagger}\,e^{i\theta}\,\sinh r$, and $\hat{S}^{\dagger}(\zeta)\,\hat{a}^{\dagger}\,\hat{S}(\zeta)=\hat{a}^{\dagger}\,\cosh r-\hat{a}\,e^{-i\theta}\,\sinh r$, the two-time correlation function is obtained as
\begin{equation}
\begin{aligned}
\bra{\zeta}\,\hat{p}_I(t_1)\,\hat{p}_I(t_2)\,\ket{\zeta}
=&
\underbrace{
\frac{m\hbar\omega}{2}\Big(\sinh^2 r\,e^{-i\omega(t_2-t_1)}+\cosh^2 r\,e^{i\omega(t_2-t_1)}\Big)
}_{F^{(\rm sq)}(t_2-t_1)}\\\\
&\underbrace{-\frac{m\hbar\omega\,\,\sinh r\cosh r}{2}\Big(e^{i\{\omega(t_2+t_1)-\theta\}}+e^{-i\{\omega(t_2+t_1)-\theta\}}\Big)}_{G(t_2+t_1)}
\end{aligned}
\end{equation}
Since the expectation value of the field $\langle \hat{p}_I(t) \rangle$ vanishes for a squeezed state, the above two-time correlation function does not contain any mean-field contribution. As a result, the two-time correlation function for the squeezed state can be decomposed into two distinct contributions, one that depends on the time difference $F^{(\mathrm{sq})}(t_2 - t_1)$ and another that depends on the time sum $G(t_1 + t_2)$. As discussed previously, contributions that depend only on the time difference can be reproduced by the two-time correlation function of a suitably modeled classical gravitational field. In contrast, the time-sum function $G(t_1 + t_2)$ does not arise for classical gravitational fields that are stationary, namely fields whose correlations depend only on the time difference. Therefore, the contribution to the detector response originating from $G(t_1 + t_2)$ reflects the influence of quantum properties of the gravitational field on the detector dynamics and cannot be reproduced within a stationary classical gravitational-field framework.

The contribution of the function $G(t_1+t_2)$ to the detector response can be analyzed by computing the detector transition probability using equation (\ref{eq:3.7}), by explicitly choosing the two-time correlation function
$\langle \hat{p}_I(t_1)\hat{p}_I(t_2)\rangle$ to be contributed solely by the term $G(t_1+t_2)$. Accordingly, the contribution to the transition probability originating from non-classical correlations can be written as
\begin{eqnarray}\label{eq:5.4}
P^{\rm (non-classical)}_{i\to f}&=&\frac{\mathcal{G}^2}{4m^2 m^2_0}\,\,\Big|\sqrt{(n_r+1)(n_r+2)}\,\delta_{n'_r,\,n_r+2}-\sqrt{n_r(n_r-1)}\,\delta_{n'_r,\,n_r-2}\Big|^2\nonumber\\
&\,\,&\times \int^t_0dt_1\,\int^t_0dt_2\,\,G(t_2+t_1)\,e^{i\,(n'_r-n_r)\,\omega_0\,(t_2-t_1)}\nonumber\\
&=&\frac{4\pi\hbar\omega G}{L^3}\,\,\Big|\sqrt{(n_r+1)(n_r+2)}\,\delta_{n'_r,\,n_r+2}-\sqrt{n_r(n_r-1)}\,\delta_{n'_r,\,n_r-2}\Big|^2\,\,\sinh r\cosh r\nonumber\\
&\,\,&\times\frac{\cos (2\omega t-\theta)}{(n'_r-n_r)^2\omega^2_0-\omega^2}\left(\sin^2\left(\frac{1}{2}\,\omega\,t\right)-\sin^2\left(\frac{1}{2}\,(n'_r-n_r)\omega_0\,t\right)\right).
\end{eqnarray}
The quantity $P^{\rm (non\text{-}classical)}_{i\to f}$ defines the contribution to the detector transition probability that originates from the non-classical correlation structure of the gravitational field. As can be seen from equation (\ref{eq:5.4}), this contribution induces a time dependence of the detector transition probability that is non-linear in time. The magnitude of this non-linear dependence is controlled by the squeezing parameter $r$ through the factor $\sinh r \cosh r$, such that larger values of $r$ lead to increasingly significant non-linear contributions to the detector transition probability. In addition, this non-classical contribution exhibits a sensitive dependence on the squeezing phase $\theta$ and on the interaction time $t$, which is mediated by the oscillatory factor $\cos(2\omega t-\theta)$. This oscillatory factor reaches its extremal values when $2\omega t-\theta = n\pi$, with n an integer, and vanishes when $2\omega t-\theta = (n+\tfrac{1}{2})\pi$. In the latter case, the contribution arising from the function $G(t_1+t_2)$ disappears, and the detector transition probability receives no contribution from non-classical correlations, independently of the transition process considered.

Unlike the resonant contributions discussed in Section \ref{sec4}, which grow linearly with the interaction time, the non-classical contribution arising from $G(t_1+t_2)$ remains bounded and oscillatory in time. Therefore, in the long-time limit this contribution does not dominate the detector response. However, for finite interaction times comparable to the oscillation periods of the system, this effect can still produce a characteristic modulation of the transition probability that is controlled by the squeezing parameter $r$. This non-classical contribution appears in both processes involving decreases and increases in the detector energy. For example, for the transition from $n_r = 2$ to $n_r' = 0$, one obtains.
\begin{eqnarray}
P^{\rm (non-classical)}_{2\to 0}=\frac{8\pi\hbar\omega G}{L^3}\,\sinh r\cosh r\,\left(\frac{\cos (2\omega t-\theta)}{4\omega^2_0+\omega^2}\right)\left(\sin^2\left(\omega_0\,t\right)-\sin^2\left(\frac{1}{2}\,\omega\,t\right)\right),
\end{eqnarray}
and for the transition from $n_r=0$ to $n'_r=2$
\begin{eqnarray}
P^{\rm (non-classical)}_{0\to 2}=\frac{8\pi\hbar\omega G}{L^3}\,\sinh r\cosh r\,\left(\frac{\cos (2\omega t-\theta)}{4\omega^2_0-\omega^2}\right)\left(\sin^2\left(\frac{1}{2}\,\omega\,t\right)-\sin^2\left(\omega_0\,t\right)\right).
\end{eqnarray}
Although the non-linear time dependence appears in both resonance conditions, the form and magnitude of the non-linearity differ for each transition process. To estimate the order of magnitude of this transition probability, we consider the prefactor of the non-classical contribution, which scales as
\begin{eqnarray}
 P^{(\text{non-classical})} \sim \frac{8\pi \hbar \omega G}{L^3}\sinh r\cosh r.
\end{eqnarray}
By considering gravitational waves with frequency $f \sim 100\,\text{Hz}$, corresponding to $\omega \sim 10^2\,\text{s}^{-1}$ (a frequency range accessible to ground-based interferometric detectors such as LIGO \cite{LIGO}), and a characteristic detector scale $L \sim 1\,{\rm m}$, we can estimate the magnitude of this non-classical contribution. In the case of primordial gravitational waves generated during inflation, the squeezing parameter can become relatively large. As a conservative estimate, we take $r \sim 5$, although in several studies \cite{Grishchuk2, Giovannini} the squeezing parameter can lie in the range $r \sim 1-100$. With these parameters, the non-classical contribution is estimated to scale as $P^{(\text{non-classical})} \sim 10^{-37}$. This estimate indicates that the contribution remains extremely small for realistic detector parameters, unless the squeezing parameter $r$ becomes sufficiently large. In a more realistic setting, additional effects such as switching functions describing the turn-on and turn-off of the interaction, as well as environmental decoherence, would further suppress the observability of this contribution. Nevertheless, the distinctive qualitative feature in the case of gravitational waves in a squeezed state is the emergence of a non-linear time dependence in the detector transition probability. Such a time dependence does not arise for stationary classical gravitational fields or for coherent quantum states, where the dominant contributions grow linearly with the interaction time.

\section{Conclusion}\label{sec6}
In this work, we investigate the response of a detector confined in a harmonic oscillator potential when interacting with classical and quantum gravitational fields. The quantum states of the gravitational field considered include coherent states, which exhibit dynamics closest to a classical description, as well as squeezed states, which
represent quantum states with stronger non-classical properties. The detector model employed follows the framework introduced by Sen \textit{et al.} \cite{Sen}.  The detector response is analyzed through transition probabilities between its energy levels, which are then compared for interactions with a classical gravitational field, a quantum gravitational field in a coherent state, and a quantum gravitational field in a squeezed state. This comparison is carried out to investigate whether there exist contributions to the detector transition probabilities that are specifically influenced by the quantum nature of the gravitational field and cannot be reproduced within the framework of a classical stationary gravitational field.

From our analysis, we find that the influence of the gravitational field on the detector transition probabilities is explicitly encoded in the two-time correlation function of the field. In the case of a classical gravitational field and a quantum gravitational field in a coherent state, the detector transition probabilities may take either the same or different forms, depending on how the classical gravitational field is modeled. If the classical gravitational field is assumed to be purely deterministic, the resulting detector transition probabilities differ from those obtained for a quantum gravitational field in a coherent state. This difference arises because the two-time correlation function of a deterministic classical field cannot reproduce the correlation structure inherent to a coherent state. Conversely, if the classical gravitational field is extended by including a stationary fluctuating component on top of its deterministic contribution, the resulting two-time correlation function can acquire a structure similar to that of a coherent state. Under these conditions, the detector transition probabilities generated by a classical gravitational field and by a quantum gravitational field in a coherent state can become indistinguishable.

A qualitatively different behavior arises when the detector interacts with a quantum gravitational field in a squeezed state. In this case, the two-time correlation function of the gravitational field contains additional components that depend on the sum of the two times. Such a correlation structure cannot be reproduced by a classical gravitational field whose two-time correlation function is stationary. As a consequence, the contribution to the detector transition probabilities originating from this time-sum dependence can be regarded as a part of the detector response that is influenced by the non-classical properties of the gravitational field. This contribution depends explicitly on the squeezing parameter and modifies the temporal behavior of the detector transition probabilities. In particular, the presence of this non-classical contribution can lead to a non-linear time dependence in the detector transition probabilities. Accordingly, the emergence of temporal non-linearity in the detector response can be interpreted as a manifestation of the influence of the quantum nature of the gravitational field on the detector dynamics. However, this non-linear contribution does not dominate the detector transition probability when the interaction time becomes long, since the contribution originating from the non-classical properties of gravitons in a squeezed state does not grow with the interaction time. In other words, this effect mainly appears for finite interaction times. An order-of-magnitude estimate indicates that the resulting transition probability is extremely small for realistic detector parameters. Furthermore, in more realistic situations additional effects such as finite interaction times, switching functions, and environmental decoherence would further suppress the observability of this contribution. Nevertheless, the presence of this non-linear time dependence still provides a distinctive qualitative feature in the detector response when interacting with a gravitational field in a squeezed state.

\section*{Conflict of Interest}
The authors declare that they have no competing financial or non-financial interests.

\section*{Funding}
Funding information – not applicable.

\section*{Data Availability}
No new data were created or analyzed in this study. Data sharing is not applicable to this article.

\section*{Acknowledgement}
F.P.Z. would like to thank PPMI FMIPA ITB and Kemendiktisaintek RI, the Ministry of Higher Education, Science, and Technology of the Republic of Indonesia, for institutional support.  A.T. would like to thank the members of the Theoretical Physics Group at Institut Teknologi Bandung for their hospitality.

\appendix

\section{Interaction Hamiltonian for Classical Gravitational Field}\label{ApenA}
In this appendix, we present a detailed derivation of the interaction Hamiltonian for the case in which the gravitational field is treated as a classical (non-quantized) field. The formulation preserves the same coupling structure as in the quantized gravitational field case, with the difference that the gravitational field is not promoted to an operator. The interaction Hamiltonian is derived by considering the action of a detector confined in a harmonic oscillator potential, as given in equation (\ref{eq:2.7}). It is then assumed that the gravitational wave propagates along a single direction (the $z$-direction) with $|\boldsymbol{k}|=\omega$, and that it possesses only a single polarization mode, namely the $+$ polarization. Under these assumptions, the action in equation (\ref{eq:2.7}) becomes
\begin{eqnarray}
S_{RD}=\int\,dt\,\,\frac{m_0}{2}\,\,\bigg(\dot{\xi}^2-\dot{h}_{+}\,\dot{\xi}\,\xi-\omega^2_0\,\xi^2\bigg).
\end{eqnarray}
The corresponding Lagrangian associated with this action can then be obtained as follows
\begin{eqnarray}\label{eq:B.2}
L=\frac{m_0}{2}\,\,\bigg(\dot{\xi}^2-\dot{h}_{+}\,\dot{\xi}\,\xi-\omega^2_0\,\xi^2\bigg).
\end{eqnarray}
The conjugate momentum corresponding to the position variable $\xi$ is
\begin{eqnarray}\label{eq:B.3}
\pi&=&\frac{\partial L}{\partial \dot{\xi}}\nonumber\\
&=&m_0\,\dot{\xi}-\frac{m_0}{2}\,\dot{h}_{+}\,\xi.
\end{eqnarray}
Using the Lagrangian given in equation (\ref{eq:B.2}) and the corresponding conjugate momentum in equation (\ref{eq:B.3}), the Hamiltonian can be obtained, up to $\mathcal{O}(\dot{h}_+)$, as follows
\begin{eqnarray}\label{eq:B.4}
H&=&\frac{\partial L}{\partial \dot{\xi}}\,\dot{\xi}-L\nonumber\\
&\simeq&\frac{\pi^2}{2\,m_0}+\frac{m_0}{2}\,\omega^2_0\,\xi^2+\frac{\dot{h}_{+}}{2}\,\pi\,\xi.
\end{eqnarray}

It can be seen that this Hamiltonian corresponds to that of a harmonic oscillator with an additional interaction term involving the gravitational wave $\left(h_{+}\right)$. Accordingly, the canonical position $\xi$ and conjugate momentum $\pi$ can be promoted to operators according to
\begin{eqnarray}
\hat{\xi}&=&\sqrt{\frac{\hbar}{2m_0\omega_0}}\Big(\hat{\chi}+\hat{\chi}^{\dagger}\Big)\,\,\,\,\,\,\,\,\,\,\,\,\,\,\,\,\,\hat{\pi}=i\sqrt{\frac{m_0\hbar\omega_0}{2}}\Big(\hat{\chi}^{\dagger}-\hat{\chi}\Big),
\end{eqnarray}
where $\hat{\chi}^{\dagger}$ and $\hat{\chi}$ are the creation and annihilation operators satisfying the canonical commutation relation $\left[\hat{\chi},\hat{\chi}^{\dagger}\right]=1$. In operator form, the Hamiltonian in equation (\ref{eq:B.4}) can be written as $\hat{H}=\hat{H}_0+\hat{H}^{\text{int}}$, with the interaction Hamiltonian given by
\begin{eqnarray}\label{eq:B.6}
\hat{H}^{\text{int}}=\frac{\dot{h}_{+}}{4}\,\Big(\hat{\pi}\,\hat{\xi}+\hat{\xi}\,\hat{\pi}\Big).
\end{eqnarray}
From this result, it is evident that the field $\dot{h}_{+}(t)$, treated as a classical gravitational field, couples directly to the detector’s degrees of freedom. Based on the form of this interaction Hamiltonian, the transition probability of the detector induced by its interaction with the classical gravitational field can be calculated using the formulation discussed in Section \ref{sec3}.

\section{Two-Time Correlations of a Classical Gravitational Field with Deterministic and Fluctuating Components}\label{ApenB}
In this appendix, we show that the two-time correlation function of a classical gravitational field $\dot{h}(t)$ can also give rise to a contribution similar to the function $F^{(\mathrm{coh})}(t_2-t_1)$ that appears in the two-point correlation function of a quantized gravitational field in a coherent state. Such a contribution can arise if the classical field $\dot{h}(t)$ is defined as a deterministic field supplemented by a stationary fluctuating component. As an illustration, we consider an effective classical field of the form
\begin{eqnarray}
\dot{h}(t)\equiv A_0\,\sin (\omega t)+\delta A \,e^{-i(\omega t-\phi)},
\end{eqnarray}
where $\phi$ is a phase parameter assumed to be uniformly distributed over the interval $0<\phi<2\pi$, while $A_0$ and $\delta A$ are taken to be real constants. Under this assumption, the mean-field value of the field is given by
\begin{eqnarray}
\langle\dot{h}(t)\rangle&=&\langle A_0\sin (\omega t)\rangle+\langle\delta A\,e^{-i(\omega t-\phi)}\rangle\nonumber\\
&=&A_0\sin (\omega t)+\frac{\delta A\,e^{-i\omega t}}{2\pi}\,\,\int^{2\pi}_0 d\phi\,\,e^{i\phi}\nonumber\\
&=&A_0\sin (\omega t),
\end{eqnarray}
where the contribution from the fluctuating component vanishes upon averaging over the phase $\phi$.

The two-time correlation function of this field can then be written as
\begin{eqnarray}
\langle \dot{h}(t_1)\dot{h}^*(t_2)\rangle &=&\langle A^2_0\sin (\omega t_1)\,\sin (\omega t_2)\rangle+\langle A_0\sin (\omega t_1)\,\,\delta A\,e^{i(\omega t_2-\phi)}\rangle+\langle\delta A\,e^{-i(\omega t_1-\phi)}\,\,A_0\sin (\omega t_2)\rangle\nonumber\\\nonumber\\
&\,\,&+\langle |\delta A|^2\,e^{-i(\omega t_1-\phi)}\,e^{i(\omega t_2-\phi)}\rangle\nonumber\\\nonumber\\
&=&\underbrace{\langle \dot{h}_I(t_1)\rangle\,\langle \dot{h}_I(t_2)\rangle}_{\text{mean field}}+\underbrace{|\delta A|^2\,e^{-i\omega( t_2-t_1)}}_{\text{difference-time correlation function}}.
\end{eqnarray}
Upon averaging over the phase $\phi$, the cross terms in the two-time correlation function vanish. This averaging effectively models a classical stochastic background with stationary correlations. As a result, the two-point correlation function consists of only two contributions, namely a mean-field term and an additional term that depends on the time difference. This correlation structure has the same form as the two-point correlation function obtained for a quantized gravitational field in a coherent state. Therefore, this example demonstrates that the two-time correlation function of a deterministic classical gravitational field, when supplemented with a stationary fluctuating component, can, in principle, reproduce the correlation structure that arises in a quantized gravitational field in a coherent state within a classical stochastic framework.

\section{Integration of the Two-Time Correlation Function and the Neglect of Cross Terms}\label{ApenC}

In this appendix, we provide a justification for neglecting the cross terms that arise in the integration of the two-time correlation function appearing in the calculation of the detector transition probability. As an illustrative example, we consider the case of a classical (non-quantized) gravitational wave. As discussed in Section \ref{subsec4.1}, the two-time correlation function for a deterministic classical field reduces to the ordinary product $\dot{h}_+(t_1)\dot{h}_+(t_2)$. Following the same assumptions as in Section \ref{subsec4.1}, if the gravitational wave is given by $h_{+}(t)=2f_0\cos(\omega t)$, then
\begin{eqnarray}
\dot h_{+}(t_1)\dot h_{+}(t_2)
&=&4\,f_0^2\,\omega^2\,\sin(\omega t_1)\sin(\omega t_2).\nonumber\\
&=&f_0^2\,\omega^2\,\,\Big(e^{-i\omega t_2}-e^{i\omega t_2}\Big)\Big(e^{i\omega t_1}-e^{-i\omega t_1}\Big).
\end{eqnarray}
Substituting this expression into equation (\ref{eq:4.2}), the integration of the two-time correlation function can be written as
\begin{align}\label{eq:C.2}
\int^t_0 dt_1\,\int^t_0 dt_2\,\,&\dot{h}_{+}(t_1)\dot{h}_{+}(t_2)\,e^{i(n'_r-n_r)\,\omega_0\,(t_2-t_1)}\nonumber\\
&=f^2_0\,\omega^2\,\,\underbrace{\int^t_0 dt_1\,\int^t_0 dt_2\,\,\Big(e^{-i\omega(t_2-t_1)}+e^{i\omega(t_2-t_1)}\Big)\,e^{i(n'_r-n_r)\,\omega_0\,(t_2-t_1)}}_{\text{Resonant terms}}\nonumber\\
&\,\,\,\,\,\,-f^2_0\,\omega^2\,\,\underbrace{\int^t_0 dt_1\,\int^t_0 dt_2\,\,\Big(e^{-i\omega(t_2+t_1)}+e^{i\omega(t_2+t_1)}\Big)\,e^{i(n'_r-n_r)\,\omega_0\,(t_2-t_1)}}_{\text{Cross terms}}.
\end{align}
Thus, the integral separates into two contributions, namely the resonant terms and the cross terms. In the following, we show that in the long-time limit the resonant contribution dominates over the cross terms, thereby justifying the approximation used in the main text.

For the resonant terms in equation (\ref{eq:C.2}), the integration takes the form
\begin{align}\label{eq:C.3}
\int^t_0 dt_1\,\int^t_0 dt_2\,\,\Big(e^{-i\omega(t_2-t_1)}&+e^{i\omega(t_2-t_1)}\Big)\,e^{i(n'_r-n_r)\,\omega_0\,(t_2-t_1)}\nonumber\\
&=\frac{4\,\sin^2\left[\,\left((n'_r-n_r)\,\omega_0-\omega\right)\,\frac{t}{2}\right]}{\left((n'_r-n_r)\,\omega_0-\omega\right)^2}+\frac{4\,\sin^2\left[\left((n'_r-n_r)\,\omega_0+\omega\right)\,\frac{t}{2}\right]}{\left((n'_r-n_r)\,\omega_0+\omega\right)^2}.
\end{align}
Using the relation
\begin{eqnarray}
\delta(y)=\frac{2}{\pi}\,\lim_{t\to\infty}\,\frac{\sin^2(\frac{y t}{2})}{y^2 t},
\end{eqnarray}
the resonant contribution can be expressed in terms of Dirac delta functions as
\begin{align}
\int^t_0 dt_1\,\int^t_0 dt_2\,\,\Big(e^{-i\omega(t_2-t_1)}+e^{i\omega(t_2-t_1)}\Big)\,&e^{i(n'_r-n_r)\,\omega_0\,(t_2-t_1)}\nonumber\\
&=2\,\pi\,t\,\delta\left((n'_r-n_r)\,\omega_0-\omega\right)+2\,\pi\,t\,\delta\left((n'_r-n_r)\,\omega_0+\omega\right)
\end{align}
This expression corresponds to the Dirac delta functions appearing in equation (\ref{eq:4.10}). 
The resonant terms therefore lead to a secular growth proportional to the interaction time, 
which reflects the resonant energy exchange between the gravitational wave and the detector.

For the cross terms in equation (\ref{eq:C.2}), the integration takes the form
\begin{align}\label{eq:C.6}
\int^t_0 dt_1\,\int^t_0 dt_2\,\,\Big(e^{-i\omega(t_2+t_1)}&+e^{i\omega(t_2+t_1)}\Big)\,e^{i(n'_r-n_r)\,\omega_0\,(t_2-t_1)}\nonumber\\
&=\frac{2\,\cos \big(2\omega t\big)-2\,\cos \big((n'_r-n_r)\omega_0\,t-\omega\,t\big)-2\,\cos \big((n'_r-n_r)\omega_0\,t+\omega\,t\big)}{(n'_r-n_r)^2\,\omega^2_0+\omega^2}.
\end{align}
Unlike the resonant contribution in equation (\ref{eq:C.3}), whose magnitude grows with the interaction time, the cross-term integral in equation (\ref{eq:C.6}) does not increase with time. Instead, its magnitude remains bounded as
\begin{eqnarray}
\bigg|\int^t_0 dt_1\,\int^t_0 dt_2\,\,\Big(e^{-i\omega(t_2+t_1)}+e^{i\omega(t_2+t_1)}\Big)\,e^{i(n'_r-n_r)\,\omega_0\,(t_2-t_1)}\bigg|\leq\frac{6}{(n'_r-n_r)^2\,\omega^2_0+\omega^2}
\end{eqnarray}
This shows that the cross-term contribution remains finite and does not grow with the interaction time. Consequently, in the long-time limit the resonant terms, which scale linearly with $t$, dominate the transition probability, whereas the rapidly oscillating cross terms remain bounded and become negligible in comparison. For this reason, the cross terms are consistently neglected in the calculations presented in Section \ref{sec4}. The same argument applies to analogous oscillatory contributions that appear in the coherent- and squeezed-state calculations.

\end{document}